%% file: main.tex
\documentclass[%
 reprint,
superscriptaddress,
showkeys, 
amsmath,amssymb,
aps,
longbibliography,
]{revtex4-1}

\usepackage{graphicx}
\usepackage{dcolumn}
\usepackage{bm}
\usepackage{natbib}
\usepackage{mathtools}
\usepackage{siunitx}
\usepackage{amsthm}
\usepackage{bbm}
\usepackage{amsmath}
\usepackage{color}

\newcommand{\beq}{\begin{equation}}
\newcommand{\eeq}{\end{equation}}
\DeclareMathOperator{\id}{\mathbbm{1}}
\definecolor{JM}{RGB}{4,116,149}

\definecolor{LB}{RGB}{134,41,198}

\definecolor{TB}{RGB}{255,154,0}

\theoremstyle{definition}
\newtheorem{definition}{Definition}[section]
\theoremstyle{remark}
\newtheorem*{remark}{Remark}
\newtheorem{lemma}{Lemma}
\newtheorem{theorem}{Theorem}

\DeclareMathOperator{\Tr}{Tr}

\begin{document}

\title{ Molecular Docking with Gaussian Boson Sampling}
\author{Leonardo Banchi}
\affiliation{Xanadu, 372 Richmond St W, Toronto, M5V 1X6, Canada}
\author{Mark Fingerhuth}%
\affiliation{ProteinQure Inc., 192 Spadina Ave, Toronto, M5T 2C2, Canada}
\author{Tomas Babej}
\affiliation{ProteinQure Inc., 192 Spadina Ave, Toronto, M5T 2C2, Canada}
\author{Christopher Ing}%
\affiliation{ProteinQure Inc., 192 Spadina Ave, Toronto, M5T 2C2, Canada}
\author{Juan Miguel Arrazola}
\affiliation{Xanadu, 372 Richmond St W, Toronto, M5V 1X6, Canada}

\begin{abstract}

Gaussian Boson Samplers are photonic quantum devices with the potential to perform tasks that are intractable for classical systems.
As with other near-term quantum technologies, an outstanding challenge is to identify specific problems of practical interest where these quantum devices can prove useful.
Here we show that Gaussian Boson Samplers can be used to predict molecular docking
configurations: the spatial orientations that molecules assume when they
bind to larger proteins. Molecular docking is a central problem for pharmaceutical
drug design, where docking configurations must be predicted for large numbers of candidate molecules. We develop a vertex-weighted binding interaction graph approach,
where the molecular docking problem is reduced to finding the maximum weighted
clique in a graph.
We show that Gaussian Boson Samplers can be programmed to sample large-weight
cliques, i.e., stable docking configurations, with high probability, even in
the presence of photon loss. We also describe how outputs from the device 
can be used to enhance the performance of classical algorithms and increase their 
success rate of finding the molecular binding pose.  
To benchmark our approach, 
we predict the binding mode of a small molecule ligand to   
the tumor necrosis factor-$\alpha$ converting enzyme, a target linked to immune system diseases and cancer.
\end{abstract}

\maketitle

\section{Introduction} \label{sec:introduction}
\input{introduction}

\bigskip

\section{Background}

Before presenting our results we provide relevant background information
on Gaussian Boson Sampling, graph theory,
and molecular docking.

\input{background_gbs}

\input{background_moldock}

\bigskip

\section{GBS for molecular docking}\label{Sec:GBS for MD}

\input{results_big}

\input{results_gbs}

\input{results_gbsplus}

\bigskip

\section{Numerical results}

\input{numerics}

\bigskip

\section{Conclusions}
\label{sec:conclusion}

\input{conclusions}

\subsection*{Acknowledgments}

We  thank  Thomas  R.  Bromley,   Nathan  Killoran, Nicol\'as Quesada, Christian Weedbrook, and Maria Schuld
for fruitful discussions.

\bibliographystyle{apsrev4-1}
\bibliography{references_moldock.bib,references_gbs.bib}
\appendix
\input{app_laplacian.tex}
\input{app_labdistgraph.tex}
\setcounter{table}{0}
\renewcommand{\thetable}{S\arabic{table}}
\input{app_data.tex}
\setcounter{figure}{0}
\renewcommand{\thefigure}{S\arabic{figure}}
\input{app_suppfig.tex}

\end{document}

%% file: introduction.tex
In his lecture ``Simulating Physics with Computers" \cite{feynman1982simulating}, Richard Feynman famously argued that classical computing techniques alone are insufficient to simulate quantum physics. Since then, significant progress has been made in formalizing this intuition by finding explicit examples of quantum systems whose classical simulation can be convincingly shown to require exponential resources. An example is Boson Sampling, first introduced by Aaronson and Arkhipov~\cite{aaronson2011computational}. In this paradigm, identical photons interfere by passing through a network of beam-splitters and phase-shifters, and are subsequently detected at the output ports of the network. Despite the simplicity of this model, it has been shown that, under standard complexity-theoretic conjectures, generating samples from the output photon distribution requires exponential time on a classical computer \cite{aaronson2011computational, clifford2018classical, neville2017classical}. Several variants of boson sampling have been proposed that aim at decreasing the technical challenges with its experimental implementation \cite{lund2014boson, bentivegna2015experimental,latmiral2016towards, hamilton2017gaussian, chabaud2017continuous, lund2017exact, hamilton2017gaussian, kruse2018detailed,vernon2018scalable}.

Most efforts in the study of Boson Sampling have been focused on its viability to disprove the Extended Church-Turing thesis \cite{wolfram1985undecidability}; not on its potential practical applications. Nevertheless, it is possible to ask: if Boson Sampling devices are powerful enough that they cannot be simulated with conventional computers, is there a way of programming them to perform a useful task? In fact, practical applications of Boson Sampling have already been reported. In Ref.~\cite{huh2015boson}, it was shown that a Boson Sampling device can be used to efficiently estimate the vibronic spectra of molecules, a problem for which in general no efficient algorithm is known. Proof-of-principle demonstrations have also been reported \cite{clements2017experimental,sparrow2018simulating}. Additionally, Refs.~\cite{arrazola2018using, arrazola2018quantum, bradler2018gaussian} discuss how a specific model known as Gaussian Boson Sampling (GBS) can be employed in combinatorial optimization problems concerned with identifying large clusters of data.


Molecular docking is a computational method for predicting the optimal interaction of two molecules, typically a small molecule ligand and a target receptor. This method works by searching the configurational space of the two molecules and scoring each pose using a potential energy function. Using molecular structures to determine stable ligand-receptor complexes is a central problem in pharmaceutical drug design \cite{de2016role,kitchen2004docking,meng2011molecular,gschwend1996molecular,shoichet2002lead}.
Several techniques for finding stable ligand-receptor configurations have been
developed, including shape-complementarity methods
\cite{shentu2008context,shoichet1992molecular,shoichet1993matching,goldman2000qsd,desjarlais1986docking,dias2008molecular}
and molecular simulation of the ligand-receptor interactions
\cite{wu2003detailed,alonso2006combining}, which vary in their computational requirements.
For high-throughput virtual screening of large chemical libraries, it is desirable to
search and score ligand-receptor configurations using as few computational resources as
possible \cite{shoichet2004virtual}.
Motivated by these computational problems, 
several recent efforts have focused on practical applications of near-term quantum computers in the life sciences~\cite{moll2017quantum,streif2018solving,o2016scalable,reiher2017elucidating,fingerhuth2018quantum,babej2018coarse,li2018quantum,hernandez2017enhancing,hernandez2016novel}.

In this work, we show how GBS can be used to solve the molecular docking
problem. 
We extend the binding interaction graph approach, where the problem of identifying docking configurations can be reduced to finding large clusters in weighted graphs \cite{CombinatorialBindingKuhl1983,DockPaper1982}.
We then show how GBS devices can be programmed to sample from distributions that assign large probabilities to these clusters, thus helping in their identification.
Docking configurations can be obtained by direct sampling or by hybrid algorithms where the GBS outputs are post-processed using classical techniques.
We apply our method through numerical simulations to find molecular docking configurations for a known ligand-receptor interaction \cite{rao2007novel}. Several therapeutic agents targeting this protein have entered into clinical trials for both cancer and inflammatory diseases \cite{Moss2008}.
\bigskip

%% file: background_gbs.tex
\subsection{Gaussian Boson Sampling}
Quantum systems such as the quantum harmonic oscillator 
or the quantized electromagnetic field can be described by phase-space methods. Here, each state is uniquely determined by a quasi-probability distribution such
as the Wigner function $W(x,p)$ over its position $x$ and momentum $p$
variables \cite{serafini2017quantum}. A quantum state is called Gaussian 
if its Wigner function is Gaussian \cite{weedbrook2012gaussian}. 
Any multi-mode Gaussian state $\rho$ is parametrized by its first and second moments, 
namely the displacement $\alpha_j = \Tr[\rho\hat\xi_j]$ and the covariance matrix $\sigma$ with entries 
$\sigma_{jk} = \Tr[\{\hat\xi_j,\hat\xi_k\}]/2$, where $\hat\xi_j$ is a vector of 
creation and annihilation operators: calling $M$ the number of modes,
$\hat\xi_j=\hat a_j=(\hat x_j+i\hat p_j)/\sqrt2$ and $\hat \xi_{M+j}=\hat a_j^\dagger$ 
for  $j=1,\dots,M$.
Gaussian quantum states are ubiquitous in quantum optics, and have enabled 
detailed theoretical modeling and coherent manipulations in experiments
\cite{weedbrook2012gaussian,vogel2006quantum}. 

In spite of their infinite-dimensional Hilbert space, Gaussian 
states can be simulated efficiently, as their evolution can be modeled by linear transformations such as Bogoliubov rotations~\cite{bartlett2002efficient}. However, when non-Gaussian measurements are employed, e.g., via photon-counting detectors \cite{lund2014boson,hamilton2017gaussian} 
or threshold detectors \cite{quesada2018gaussian}, modelling measurement outcomes 
becomes extremely challenging even for supercomputers. 
Indeed, it has been shown that under standard complexity assumptions, sampling from the resulting probability distribution cannot be done
in polynomial time using classical resources \cite{aaronson2011computational, hamilton2017gaussian,kruse2018detailed}. 

For a Gaussian state with zero displacement and covariance matrix $\sigma$, the 
Gaussian Boson Sampling (GBS) distribution obtained by measuring the state with photon-counting detectors is given by \cite{hamilton2017gaussian}:
\begin{equation}
	p(S)= \frac{\rm Haf(\mathcal A_S)}{n_1!\dots n_M! 
	\;\sqrt{\det(\sigma+\id/2)}}~,
	\label{e:GBS}
\end{equation}
where 
$ 
	\mathcal A = \begin{pmatrix} 0 & \id \\\id & 0 \end{pmatrix}
	\left[\id - (\sigma + \id/2)^{-1}\right]~,
$ 
and $\mathcal A_S$ is a $2N\times 2N$ submatrix of $\mathcal{A}$, with $N=\sum_{j=1}^M n_j$. 
The set $S=(n_1,\dots,n_M)$ defines a measurement outcome, where $n_j$ is the 
number of photons in mode $j$, and the submatrix $\mathcal A_S$ is obtained by selecting rows and columns of $\mathcal A$, as described in 
Ref.~\cite{hamilton2017gaussian}. The function ${\rm Haf}(\mathcal A_S)$ is the 
Hafnian of $\mathcal A_S$, a matrix function which is \#P-Hard to approximate for worst-case instances \cite{caianiello1953quantum, barvinok2016combinatorics, bjorklund2018faster}. 
For a $2N\times 2N$ matrix $A$, it is defined as
\begin{equation}
{\rm Haf}(A) = \sum_{\mathcal M\in {\rm PMP}} \prod_{(i,j)\in \mathcal M} A_{ij},
\end{equation}
where $\rm PMP$ is the set of perfect matching permutations, namely 
the possible ways of partitioning the set ${1,\dots,2N}$ into subsets of size 2.
When threshold detectors are employed \cite{quesada2018gaussian}, the output 
is a binary variable $s_j$ for each mode: $s_j=1$ corresponds to a ``click'' from 
the $j$th detector that occurs whenever $n_j>0$; on the other hand, $s_j=0$ for 
$n_j=0$. The probability distribution with threshold detectors can be obtained by 
summing infinitely many probabilities from Eq.~\eqref{e:GBS} or via closed-form 
expressions that require evaluating an exponential number of matrix determinants \cite{quesada2018gaussian}.

\subsection{GBS to find dense subgraphs  }
When $A$ is the adjacency matrix of an unweighted graph $G$, the Hafnian of $A$ is equal to the number of perfect matchings in $G$. 
Using mathematical properties of the Hafnian, it was shown in Ref.~\cite{bradler2018gaussian} that 
a GBS device can be programmed to sample from a distribution $p(S)\propto \frac{|{\rm Haf}(A_S)|^2}{c^N}$.
The parameter $c$ depends on the spectral properties of $A$ and can be tuned to 
lower the probability of observing photon collisions, i.e.,  $n_j\ge 2$ for some $j$.
More details are provided in Appendix \ref{a:technical}. 
In the collision-free subspace, $A_S$ is the adjacency matrix of the subgraph specified 
by the vertices $j$ for which $n_j=1$, and ${\rm Haf}(A_S)$ is equal to 
the number of perfect matchings in this subgraph. Therefore, a GBS device can be programmed to sample large-Hafnian subgraphs with high probability. 

The density of a graph $G$ is defined as the number of 
edges in $G$ divided by the number of edges of the complete graph. 
Intuitively, a subgraph with a high number of perfect matchings should have a large density; a connection that was made rigorous in Ref.~\cite{aaghabali2015upper}. This fact was used in Ref.~\cite{arrazola2018using} 
to show that GBS devices can be programmed to sample dense subgraphs with high probability. Hybrid quantum-classical optimization algorithms can be built by combining GBS random sampling with stochastic optimization algorithms for dense subgraph identification. 

\bigskip

%% file: background_moldock.tex
\subsection{Molecular docking}
\label{sec:background_moldock}

Molecular docking is a computational tool for rational structure-based drug discovery.
Docking algorithms predict non-covalent interactions between a drug molecule (ligand) and a target macromolecule (receptor) starting from unbound three-dimensional structures of both components.
The output of such algorithms are predicted three-dimensional orientations of the ligand with respect to the receptor binding site and the respective score for each orientation.
Reliable determination of the most probable ligand orientation, and its ranking within a series of compounds, requires accurate scoring functions and efficient search algorithms \cite{DockingMethodsReview2002}.
The scoring function contains a collection of physical or empirical parameters that are sufficient to score binding orientation and interactions in agreement with experimentally determined data on active and inactive ligands.
The search algorithm describes an optimization approach that can be used to obtain the minimum of a scoring function, typically by scanning across translational and rotational degrees of freedom of the ligand in the chemical environment of the receptor.
In the simplest case, both the ligand and the receptor can be approximated as rigid bodies, but more accurate methods can account for inherent flexibility of the ligand and receptor \cite{dias2008molecular}.
As is the case for most molecular modelling approaches, a trade-off exists between accuracy and speed.

High-performance algorithms enable molecular docking to be used for screening large compound libraries against one or more protein targets.
Molecular docking and structure-based virtual screening are routinely used in pharmaceutical research and development \cite{Kalyaanamoorthy2011}.
However, evaluating billions of compounds requires accurate and computationally efficient algorithms for binding pose prediction. Widely used approaches for molecular docking employ heuristic search methods (simulated annealing \cite{yue1990distance} and evolutionary algorithms \cite{AutoDockVina2009}) and deterministic methods \cite{FlexXDockingAlgorithm1996}.
In one combinatorial formulation of the binding problem utilized in the DOCK 4.0 and FLOG \cite{ewing2001dock,Miller1994}, an isomorphous subgraph matching method is utilized to generate ligand orientations in the binding site \cite{DockPaper1982,CombinatorialBindingKuhl1983,ewing1997dock}.
In this formulation of the binding problem, both the ligand and the binding site of the receptor are represented as complete graphs. The vertices of these graphs are points that define molecular geometry and edges capture the Euclidean distance between these points.
In order to strike a balance between the expressiveness of the graph and its size, we reduce the all-atom molecular models of the ligand and receptor to a \emph{pharmacophore} representation \cite{guner2000pharmacophore,yang2010pharmacophore}.

\begin{figure}[]
    \centering \includegraphics[width=0.49\textwidth]{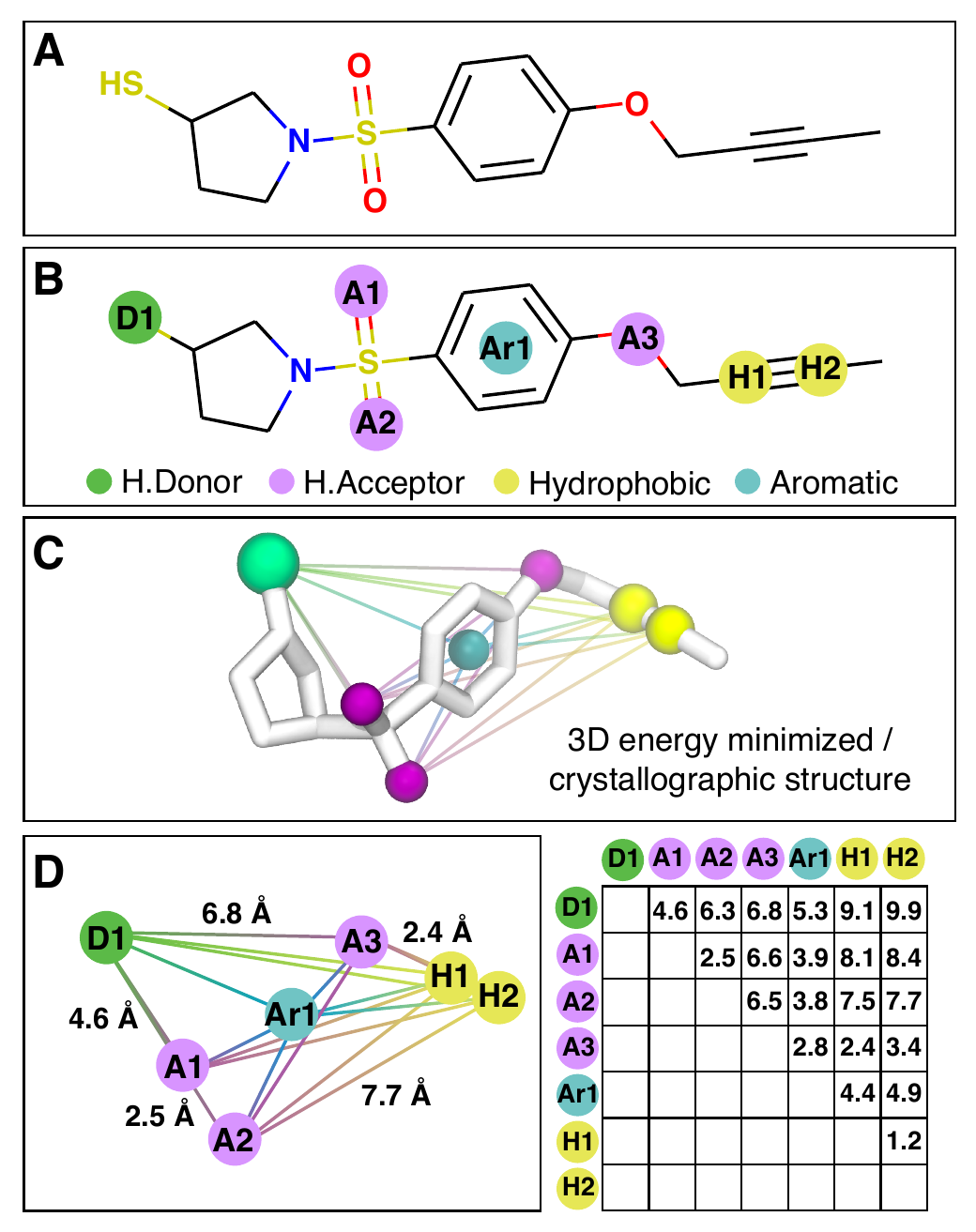}
	\caption{
        \label{fig:labelled_distance_graph}
        {\bf Construction of the labeled distance graph for a ligand molecule}.
        Panel A shows the planar structure of the ligand molecule.
        Pharmacophore points of the molecule (Panel B) are identified and their pairwise distance is measured using the known three-dimensional structure (Panel C).
        This information is combined in the labeled distance graph for the ligand molecule (Panel D), where vertices represent the pharmacophore points and edge weights their respective pairwise distance (the complete weight matrix is on the right of Panel D).}
\end{figure}

A pharmacophore is a set of points which have a large influence on the molecule's pharmacological and biological interactions. These points may define a common subset of features, such as charged chemical groups or hydrophobic regions, that may be shared across a larger group of active compounds. For the purposes of this study, we define six different types of pharmacophore points: negative/positive charge, hydrogen-bond donor/acceptor, hydrophobe, and aromatic ring.
In the graph representation, the type of the pharmacophore point is preserved as a label associated with its vertex.
Hence we refer to this molecular graph representation as a \textit{labeled distance graph}
(see also Appendix~\ref{a:graphrepr}).
As illustrated in Fig.~\ref{fig:labelled_distance_graph}, a labeled distance graph is 
constructed as follows for both the ligand and receptor:
\begin{enumerate}
\item Heuristically identify pharmacophore points likely to be 
	involved in the binding interaction. These form the vertices of the graph.
\item Add an edge between every pair of vertices and set its weight to the
	Euclidean distance between the pharmacophore points they represent.
\item Assign a label to every vertex according to the respective type of pharmacophore 
	point it represents.
\end{enumerate}



%% file: results_big.tex
\subsection{Mapping molecular docking to maximum weighted clique}

The labeled distance graphs described in Section \ref{sec:background_moldock} capture the geometric three-dimensional shapes and the molecular features of both the protein binding site and the ligand that interacts with it.
In this section, akin to \cite{CombinatorialBindingKuhl1983}, we combine these two graphs into a single \textit{binding interaction graph}.
Subsequently, we reduce the molecular docking problem to the problem of finding the maximum weighted clique.

\begin{figure}[] \centering \includegraphics[width=0.49\textwidth]{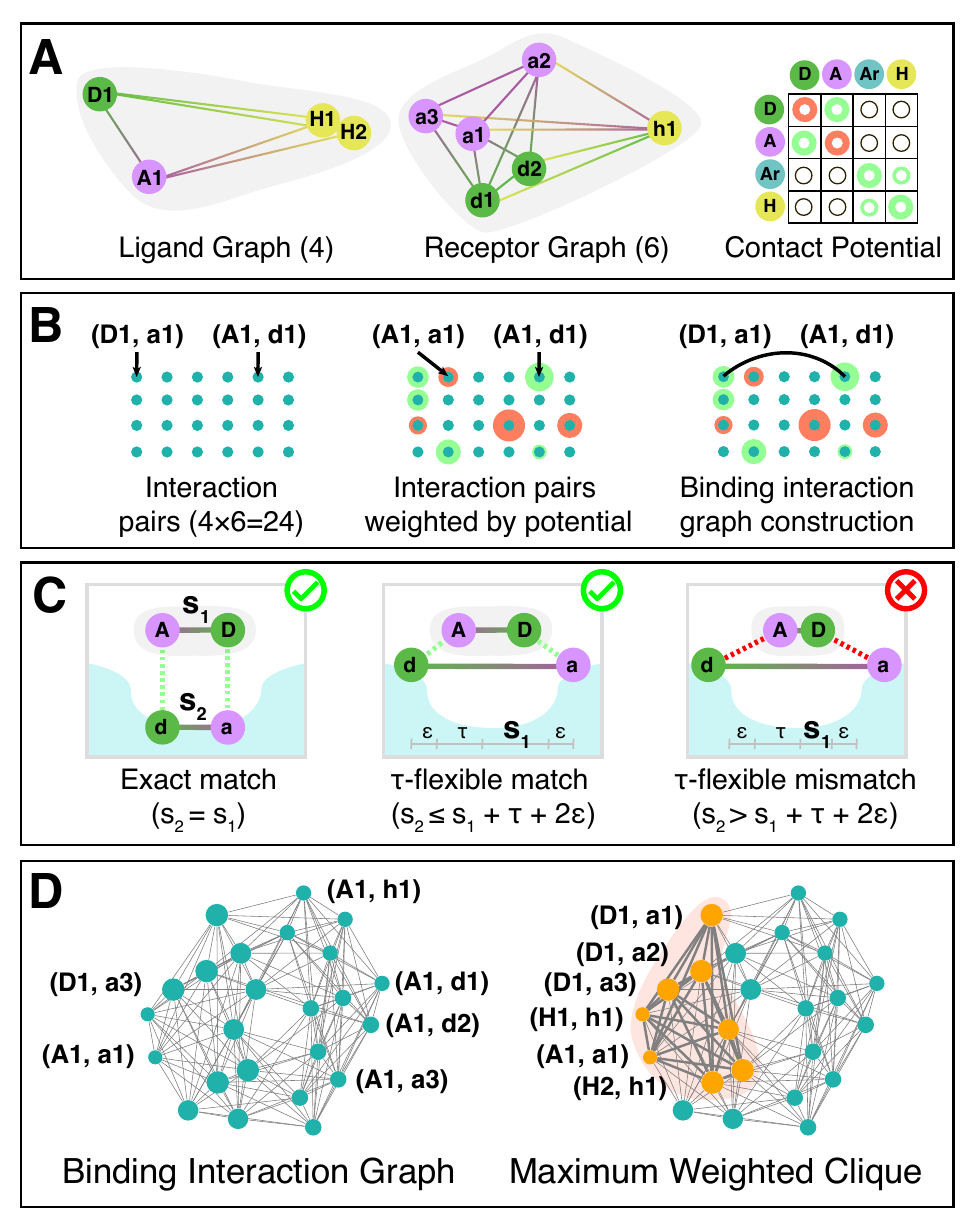}
	\caption{
        \label{fig:binding_interaction_graph}
	    {\bf Construction of the binding interaction graph}.
        Panel A depicts the inputs for the construction of the binding interaction graph -- two labeled graphs (one for the ligand and one for the receptor) and corresponding contact potential that captures the interaction strength between different types of vertex labels.
        We denote vertices on the ligand and receptor with upper and lower case letters respectively.
        The binding interaction graph is constructed (Panel B) by creating a vertex for each possible contact between ligand and the receptor, weighted by the contact potential.
        Pairs of vertices that represent compatible contacts (see Panel C for various scenarios) are connected by an edge.
        The resulting graph is then used to search for potential binding poses (Panel D).
        These are represented as complete subgraphs -- also called cliques -- of the graph, as they form a set of pairwise compatible contacts.
        The heaviest vertex-weighted cliques represent the most likely binding poses (maximum vertex-weighted clique depicted in red).
    }
\end{figure}

If two pharmacophore points are interacting, they form a \textit{contact}.
A binding pose can be defined by a set of three or more contacts that are not colinear.
We model contacts as pairs of interacting vertices of the labeled distance graphs of the ligand and the binding site. 
Consider the labeled distance graph $G_L$ of the ligand and the labeled distance graph $G_B$ of the binding site, with their vertex sets $V_L$ and $V_B$ respectively.
A \textit{contact} is then represented by a singe vertex $c_i \in V_L \times V_B$.
The set of possible contacts forms the vertices of the binding interaction graph.
In principle, any pharmacophore point of the ligand could be interacting with any pharmacophore point of the binding site, and therefore we have to consider every possible pair of corresponding interacting vertices.
Hence the number of vertices of the binding interaction graph is $nm$, where $n$ is the number of vertices of the labeled distance graph $G_L$ and $m$ is the number of vertices of the labeled distance graph $G_B$.

The goal of the binding interaction graph is to model possible binding poses via sets of contacts.
However, not every combination of contacts is physically realizable.  
Two contacts 
are not be compatible if their mutual realization would violate the geometrical shapes of the ligand and the binding site.
To model this, 
the binding interaction graph contains an edge between two contacts if and only if they are compatible.
As a result, a pairwise compatible set of contacts, i.e., such as would arise from a true binding pose, forms a complete subgraph of the binding interaction graph.
A complete subgraph, also called a {\it clique}, in a graph $G$ is a subgraph where all possible pairs of vertices are connected by an edge.

The compatibility of contacts is captured by the notion of $\tau$ flexibility
, which is illustrated in Fig.~\ref{fig:binding_interaction_graph}
(see also Appendix~\ref{a:graphrepr}). 
Even though both the ligand and the binding site can exhibit a certain amount of flexibility, in general, geometric distances between two contacts have to be approximately the same both on the ligand and the binding site.
Two contacts $(v_{l_1}, v_{b_1})$ and $(v_{l_2}, v_{b_2})$ form a \textit{$\tau$ flexible{} contact pair} if the distance between the pharmacophore points on the ligand (points corresponding to vertices $v_{l_1}$ and $v_{l_2}$) and the distance between the pharmacophore points on the binding site (points corresponding to vertices $v_{b_1}$ and $v_{b_2}$) does not differ by more than $\tau + 2\epsilon$ (see Panel C in Fig.~\ref{fig:binding_interaction_graph}).
The constants $\tau$ and $\epsilon$ describe the flexibility constant and interaction distance respectively.


In order to model varying interaction strengths between different types of pharmacophore points, 
we associate a different weight to every vertex of the binding interaction graph.
The weights are derived using the pharmacophore labels that are captured in the labeled distance graphs of the ligand and the binding site.
Given a set of labels $\mathbb{L}$, a potential function $\kappa: \mathbb{L}\times \mathbb{L} \rightarrow \mathbb{R}$ is applied to compute the weights of the individual vertices.
This allows us to bias the algorithm towards stronger intermolecular interactions.
Potential functions can be derived in several ways, ranging from pure data-based approaches such as statistical or knowledge-based potentials \cite{poole2006knowledge,mintseris2007integrating,gohlke2001statistical} to quantum-mechanical potentials \cite{kitchen2004docking}.
Details of the potential used in this study are described in Section \ref{sec:numerics}.

Hence under the model derived in this study, the most likely binding poses correspond to vertex-heaviest cliques in the binding interaction graph.
The problem of finding a maximum weighted clique is a generalization of the maximum clique problem of finding the clique with the maximum number of vertices.
When $G$ has $n$ vertices, the number of possible subgraphs is $\mathcal O(2^n)$, so a brute force approach becomes rapidly infeasible for growing values of $n$.
The max-clique decision problem is NP-hard \cite{karp1972reducibility}: as such, unless P=NP, in the worst case any exact algorithm run for superpolynomial time before finding the solution.
There are deterministic and stochastic classical algorithms for finding both the maximum cliques and maximum weighted cliques, or for finding good approximations when $n$ is large \cite{wu2015review}.

%% file: results_gbs.tex
\subsection{Max weighted clique from GBS}
In this section, we show that a GBS device can be programmed to sample from a distribution that outputs the  
max-weighted clique with high probability. The main technical challenge is to program a GBS device to sample, with high probability, subgraphs with a large total weight that are as close 
as possible to a clique. 
Consider the graph Laplacian $L=D-A$, where $D$ is the degree matrix and $A$ the adjacency matrix. The normalized Laplacian \cite{chung1997spectral}
$\tilde L = D^{-1/2}LD^{-1/2}$ is 
positive semidefinite and its spectrum is contained in $[0,2]$. More generally, we define 
a rescaled matrix 
\begin{equation}
	B = \Omega(D-A)\Omega,
	\label{e:Bmat}
\end{equation}
where $\Omega$ is a suitable diagonal matrix. If the largest entry of $\Omega$ is bounded 
as shown in Appendix \ref{a:technical}, then the spectrum of $B$ is contained in $[0,c]$, 
where $c\le 1$ can be tuned depending on 
the maximum amount of squeezing obtainable experimentally. Using the decoupling theorem
from Appendix~\ref{a:technical},
we find that a GBS device can be programmed to sample from the distribution
\begin{equation}
	p(S) \propto {[\det(\Omega_S) \; {\rm Haf}(A_S)]^2},
	\label{e:DetHaf}
\end{equation}
where we consider outputs $S=(n_1,\ldots, n_M)$ with $n_j\leq 1$ and 
$N=\sum_j n_j$ total photons. 
In the collision-free subspace, the dependence on the diagonal matrix $D$ 
disappears so we may focus on programming GBS with a rescaled adjacency 
matrix $\Omega A \Omega$. From a GBS sample $S$, we construct the subgraph $H$ of 
$G$ made by vertices $j$ with $n_j=1$. The matrix $A_S$ is the $N\times N$ adjacency matrix of 
$H$.  The Hafnian of an adjacency matrix is maximum for the complete graph, namely when 
$H$ is a clique. Therefore, for a fixed total number of photons $N$, the Hafnian term 
maximizes the probability of detecting photon configurations that correspond to a clique. 

Different choices are possible for the weighting matrix $\Omega$. For an
unweighted graph, convenient choices are either a constant $\Omega$ or
$\Omega\propto D$.  In the former case, $\det\Omega_{S} = c^{N}$ for
$c<1$, so the parameter $c$ can be tuned via squeezing in order to penalize
larger $N$, i.e., larger subgraphs (see Appendix~\ref{a:bias}).
In the latter case, $\det\Omega = c^{N}
\det D$ is proportional to the Narumi-Katayama index \cite{gutman2012some},
 which describes some topological properties of the graph.  Similarly to
the Hafnian, it is maximum when $H$ is a clique.  

For a vertex-weighted graph,
we can use the freedom of choosing $\Omega$ to favour 
subgraphs with larger total weight. 
There are multiple ways of introducing the weights $w_j$ in $\Omega$ and a convenient 
choice is 
\begin{equation}
	\Omega_{ii} = c(1+\alpha  w_i),
	\label{e:weightrule}
\end{equation}
where $c$ is a normalization to ensure the correct spectral properties and $\alpha>0$ 
is a constant. When $\alpha$ is small, the determinant term  $\det \Omega_S
\approx 1 + \alpha \sum_{j:n_j=1} w_j$ is large when the subgraph $H$ has a large  
total weight. This is useful for the max-weighted clique problem as it 
introduces a useful bias in the GBS probability of Eq.~\eqref{e:DetHaf} that favours heavier subgraphs. 
However, if $\alpha$ is too large, the Hafnian term in Eq.~\eqref{e:DetHaf} becomes 
less important and GBS will sample heavy subgraphs that typically do not 
contain cliques. To prevent this occurrence, the parameter $\alpha$ must be chosen 
carefully. Ideally, the weights should give a positive bias to heavy cliques, but 
should not favour heavy subgraphs that are not cliques.
More details are discussed in Appendix \ref{a:technical}.

%% file: results_gbsplus.tex
\subsection{Hybrid algorithms}
\label{s:gbsclassical}

\begin{figure}[t]
	\centering
	\includegraphics[width=0.9\linewidth]{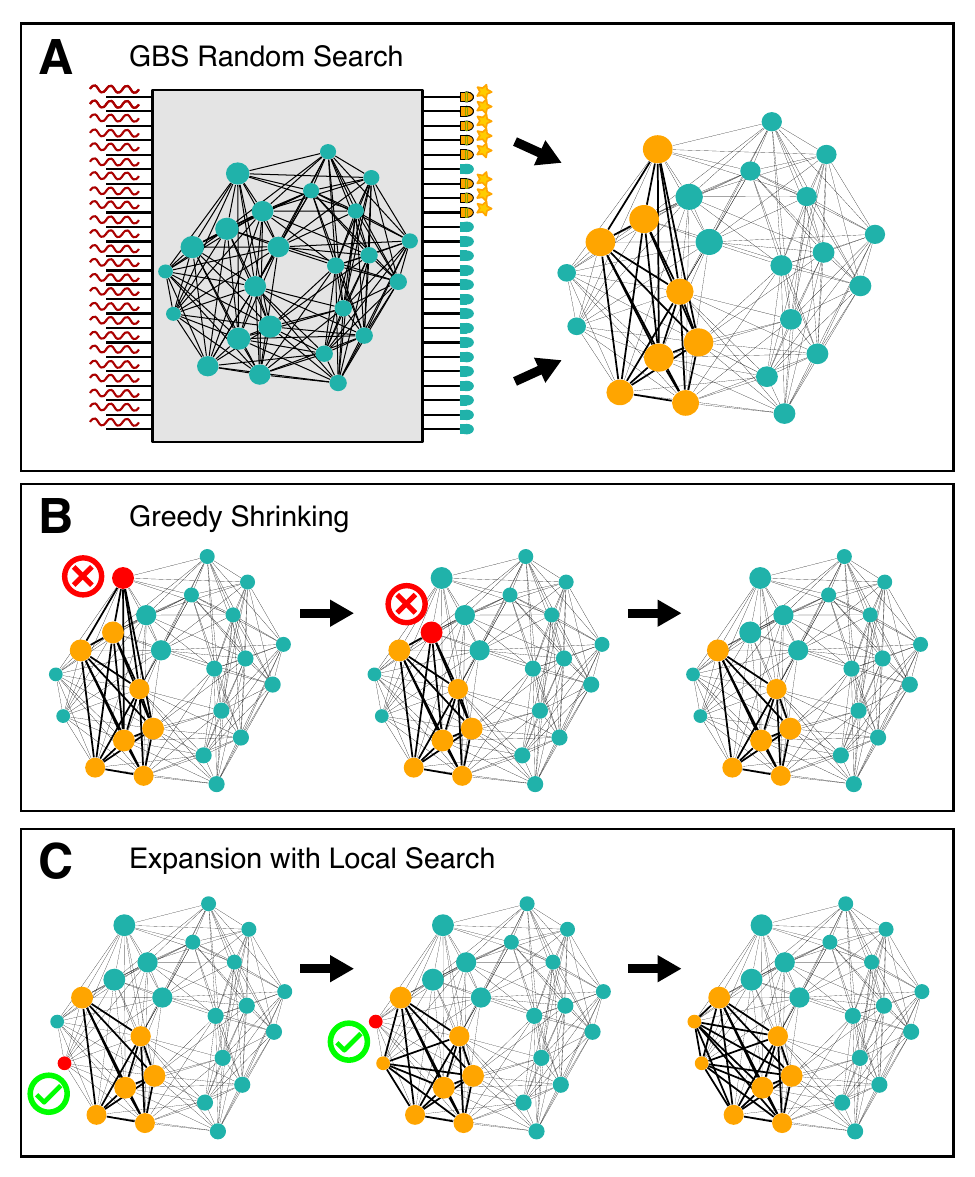}
	\caption{ {\bf Schematics of the protocol.} 
		Squeezed light is injected from the left into a GBS device, 
		which is programmed to sample from a vertex-weighted graph. Presence (star) or absence of photons 
		is measured by detectors on the right. GBS Random Search (panel A):
		based on the ports where photons have been detected, we construct a subgraph 
		(yellow vertices, dark edges) and check if it is a clique. 
		If it is not a clique, greedy shrinking (panel B) iteratively removes a vertex (red node with a cross) until 
		the remaining ones form a clique. Two shrinking iterations are shown in (B) from 
		left to right. In local search~(panel C), the found clique is expanded
		by iteratively adding, as long as possible, 
		a neighbouring vertex (red node with a tick) to get a bigger clique. 
	}
	\label{fig:drawing}
\end{figure}

GBS devices can in principle have a very high sampling 
rate -- primarily limited by detector dead time -- so just by observing the photon 
distribution it is possible to extract the maximum weighted clique for 
small enough graphs. We call this simple strategy {\it GBS random search} 
-- see Fig.~\ref{fig:drawing} for a graphical explanation of the method. 
However, selecting photon outcomes that 
correspond only to cliques means wasting samples that are potentially 
close to the solution. Indeed, an optimally programmed GBS device will sample from 
both the correct solution and neighboring configurations with high probability. 
Therefore, we propose two algorithms to post-process all GBS data which incur an overhead in run time but are especially useful for finding cliques in larger graphs.

{\it Greedy Shrinking:} 
Starting from an output subgraph $H$ from GBS, vertices are removed
based on a local rule until a clique is found
-- see Fig.~\ref{fig:drawing} for a graphical explanation of the method. 
Removal is based on vertex degree and weight. Vertices with small degree are unlikely to be 
part of a clique making them good candidates to be discarded. The role of the weights is less straightforward:
vertices with low weight may not be part of the max-weighted
clique, but this assumption may be incorrect if the clique is made by a 
heavy core together with a few light vertices. Because of this, vertex degree
is prioritized over vertex weight during the greedy shrinking stage. More precisely, the algorithm proceeds as follows: 
\begin{enumerate}
\item From a GBS outcome, build a subgraph $H$ with vertices corresponding
to the detectors that ``click''. 
\item If $H$ is a clique, return $H$.
\item Otherwise, set $v$ as the set of vertices in $H$ with smallest degree.
\item Set $w$ as the subset of $v$ with lowest weight.
\item Remove a random element of $w$ from $H$ and go back to step 2.
\end{enumerate} 

{\it Expansion with Local Search:} GBS provides high-rate samples from max-cliques, and greedy shrinking enhances 
the probability of finding a solution via classical post-processing of sampled 
configurations. 
We may increase the probability of finding the solution even further, at the cost of 
a few more classical steps. This is done by employing a local search algorithm that tries
to expand the clique with neighbouring vertices, as shown also in Fig.~\ref{fig:drawing}. 
Algorithms such as Dynamic Local Search (DLS) \cite{pullan2006dynamic}
and Phased Local Search (PLS) \cite{pullan2006phased}
are among the best-performing classical algorithms for max-clique \cite{wu2015review}. 
These algorithms usually start with a candidate clique 
formed by a single random vertex, and then try to expand the clique size and replace 
some of its vertices by locally exploring the neighbourhood. 
More precisely, the following iteration is repeated until a sufficiently
good solution is found, or the maximal number of steps is reached:
\begin{enumerate}
	\item {\it Grow stage}: Starting from a given clique, generate the set of
		vertices that are connected to all vertices in the clique. If this set is
		non-empty, select one vertex at random, possibly with large weight, and add it to the clique. 
	\item {\it Swap stage}: If the above set is empty, generate the
		set of vertices that are connected to all vertices in the clique except
		one (say $v$). From this new set, select a vertex at random and swap it with $v$. 
		This gives a new
		clique of the same size but with different vertices, thus constituting a
		local change to the clique. For max-weighted clique, the swapping rule also 
		considers vertex weight. 
\end{enumerate}
An important aspect of the above local search is that, 
at each iteration step, the candidate solution is always a clique and the algorithm tries 
to expand it as much as possible. 
GBS can be included in this strategy in view of its ability to 
provide a starting configuration that is not a mere random vertex. 
Indeed, a GBS output after greedy shrinking is always a clique, with a  
comparatively large  probability of being close to the maximum clique. 
In case the candidate output from greedy shrinking is not the maximum clique, then 
it can be expanded with a few iterations of local search. 
Since the cliques sampled from a carefully 
programmed GBS device are, with high probability, larger than just a random vertex,
the number of classical expansion steps is expected to be significantly reduced. 
This will be demonstrated with relevant numerical examples in the following section.

%% file: numerics.tex
\label{sec:numerics}
We study the binding interaction between the tumor necrosis factor-$\alpha$ converting enzyme (TACE) and a thiol-containing aryl sulfonamide compound (AS).
TACE was chosen due to the planar geometry of the active site cleft and its high relevance to the pharmaceutical industry.
Due to its role in the release of membrane-anchored cytokines like the tumor necrosis factor-$\alpha$, it is a promising drug target for the treatment of certain types of cancer, Crohn's disease and rheumatoid arthritis \cite{Murphy2008,moss2008drug,rosejohn2013}.
The ligand under consideration is part of a series of thiol-containing aryl sulfonamides which exhibit potent inhibition of TACE, and is supported by a crystallographic structure \cite{rao2007novel}. This complex provides an important testbed to benchmark our GBS-enhanced method. 
As we will show, our method is able to find the correct binding pose without requiring all-atom representation or simulation of the ligand/receptor complex.
	
The binding interaction graph for the TACE-AS complex is constructed by first extracting
all the pharmacophore points on ligand and receptor using the software package
\texttt{rdkit} \cite{rdkit}. To simplify numerical simulations, we identity
the relavant pairs of pharmacophore points on the ligand and receptor
that are within a distance of $4$\r{A} of each other, and whose
label pairs are either hydrogen donor/acceptor, hydrophobe/hydrophobe, negative/positive charge, aromatic/aromatic.
After this procedure, we get 4 points on the ligand and 6 points on the receptor and create two labelled distance graphs as illustrated in Fig.~\ref{fig:labelled_distance_graph}.
The knowledge-based potential is derived by combining information from
PDBbind \cite{liu2014pdb,wang2005pdbbind,wang2004pdbbind},
a curated dataset of protein-ligand interactions,
and the Drugscore potential \cite{gohlke2000predicting,gohlke2000knowledge,mooij2005general}.
More details are presented in Appendix~\ref{a:data}, where
the resulting knowledge-based potential is shown in Table~\ref{t:potential}.

Using this knowledge-based potential, we combine the two labelled distance graphs into the TACE-AS binding interaction graph as shown in Fig.~\ref{fig:binding_interaction_graph}. 
A summary of our graph-based molecular docking approach is shown in Fig.~\ref{fig:graph_workflow_overview}, which includes a molecular rendering of the predicted binding interactions of the AS ligand in the TACE binding site using the crystallographic structure of this complex (PDB: 2OI0) \ref{rao2007novel}.
These interactions correspond to the maximum vertex-weighted clique in the TACE-AS graph. This set of pharmacophore interactions can be used as constraints in a subsequent round of molecular docking to deduce three-dimensional structures of the ligand-receptor complex \ref{Hindle2002,verdonk2004}.
We now study the search of the maximum weighted clique on the TACE-AS graph via a hierarchy of algorithms in increasing order of sophistication. As discussed previously, these are:

\begin{figure}[]
    \centering \includegraphics[width=0.485\textwidth]{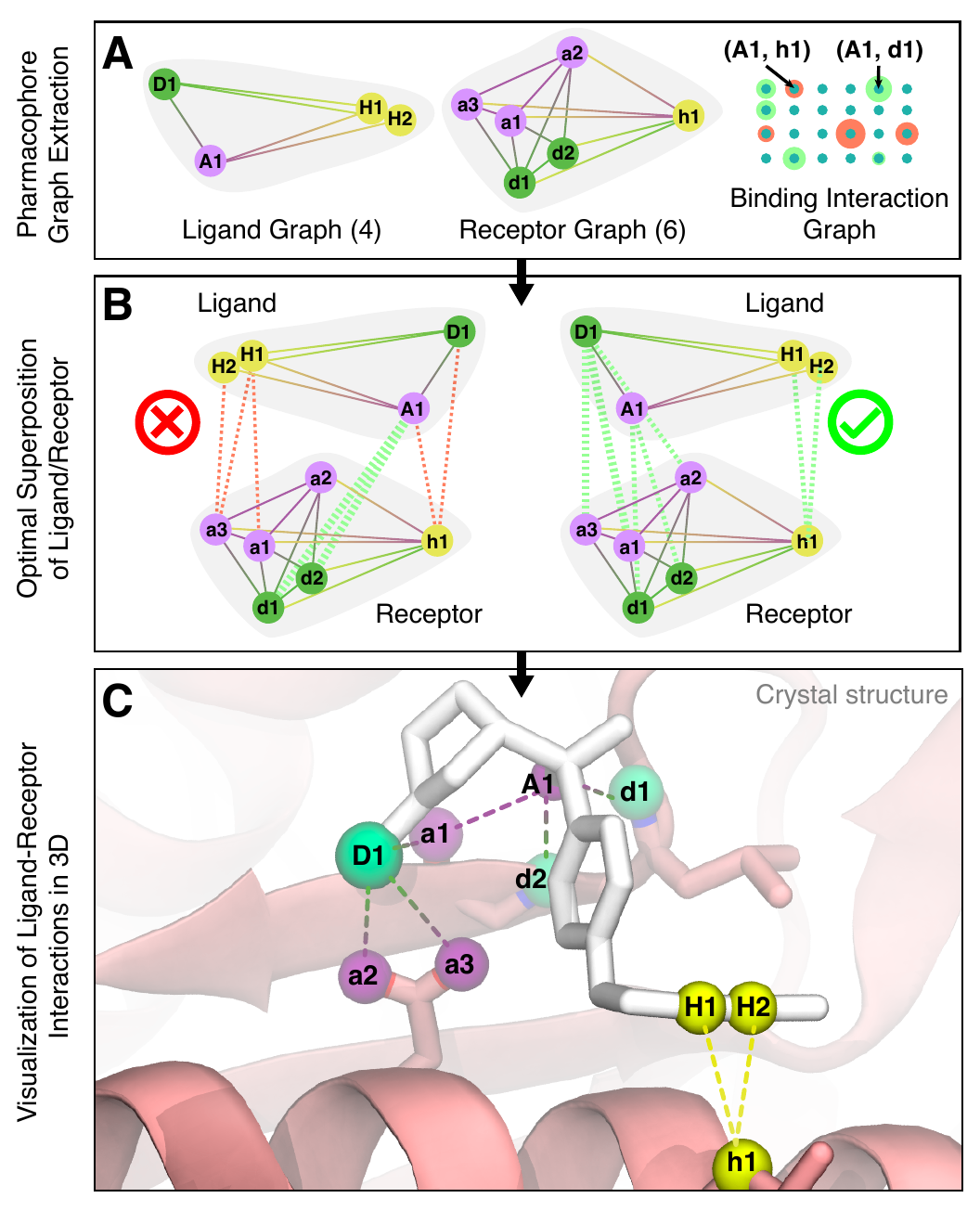}
	\caption{
        \label{fig:graph_workflow_overview}
	    {\bf Graph-based molecular docking of an aryl sulfonamide compound to TACE}.
        Panel A shows the two labelled distance graphs -- one for the aryl sulfonamide compound and one for the TACE receptor -- and the resulting TACE-AS binding interaction graph.
        Construction of the labelled distance graph and binding interaction graph are described in Figs.~\ref{fig:labelled_distance_graph} and \ref{fig:binding_interaction_graph}.
        Pharmacophore points on the ligand and receptor are labelled with upper- and lowercase letters, respectively.
        The search for the maximum vertex-weighted clique within the TACE-AS graph is illustrated in Panel B.
        Each clique in the TACE-AS graph correspond to a different superposition of the ligand molecule and the TACE receptor.
        The correct ligand-receptor superposition corresponding to the maximum weighted clique in the TACE-AS graph is shown on the right.
        Panel C visualizes the crystallographic structure of the TACE-AS complex with optimal ligand-receptor interactions correctly predicted by the maximum weighted clique. We omit the metal cofactor in the enzyme active site for visual clarity, as it was not considered as a pharmacophore point under our procedure.
    }
\end{figure}

\begin{enumerate}
	\item {\it Random search}: Generate subgraphs at random and pick the cliques with
		the largest weight among the outputs.
	\item {\it Greedy shrinking}: Generate a large random subgraph and remove vertices
		until a clique is obtained. Vertices are removed by taking into account
		both their degree and their weight.
	\item {\it Shrinking + local search}: Use the output of the greedy shrinking algorithm as the input to a local search algorithm.
\end{enumerate}
These form a hierarchy in the sense that random search is a subroutine of greedy shrinking,
which is itself a subroutine of shrinking + local search.
For each of these algorithms we compare the performance of standard classical strategies
with their quantum-classical hybrid versions introduced in Sec.~\ref{s:gbsclassical}, where the random
subgraph is sampled via GBS.
For a fair comparison with GBS-based approaches, the classical data is generated as follows: we first sample a subgraph size $N$ from a normal distribution with the same mean $\langle N\rangle$ and variance $\Delta N^2$ as the GBS distribution, then uniformly generate a random subgraph with size $N$.

We begin our analysis with a pure GBS random search.
We consider GBS with threshold detectors, which register measurement outcomes as either `no-click' (absence of photons) or `click' (presence of one or more photons). We employ either a brute force approach
to calculate the resulting probability distribution or, when that becomes infeasible,
the exact sampling algorithm discussed in Refs.~\cite{quesada2018gaussian,gupt2018classical}.
Given the complexity of simulating GBS with classical computers, for simplicity in numerical benchmarking,
we first consider the simpler case where the maximum clique size is known, so we can post-select
GBS data to have a fixed number of detection clicks. This drastically simplifies
numerical simulations (see Appendix~\ref{a:postselection} for details),
at the expense of disregarding data that would otherwise be present in an experimental setting.

\begin{figure}[t]
	\centering
	\includegraphics[width=0.9\linewidth]{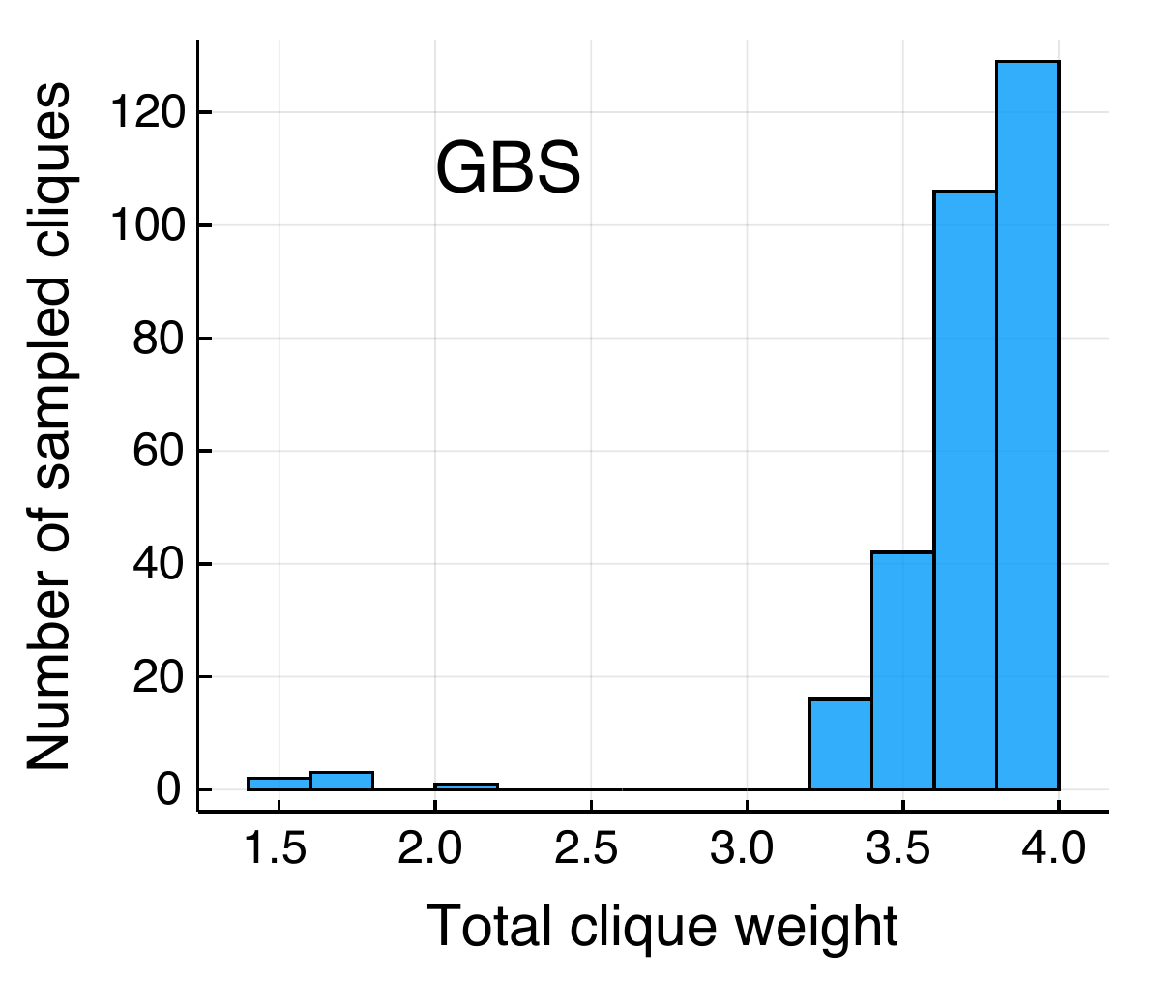}
	\caption{ {\bf GBS random search sampling rate}.
		Number of cliques sampled from a GBS device as a function of the total
		clique weight $\sum_{j\in C} w_j$.
		GBS output has been post-selected to $10^5$ samples with total number of
	  detector clicks $N=8$. With the same sample size, classical random search only
		found three cliques (not shown), none of them with maximum weight.
	}
	\label{fig:gbs_rand_search}
\end{figure}

For the TACE-AS binding interaction graph,
the largest and heaviest cliques both have eight vertices, so we fix
$N=8$. There are a total of 19 cliques of this size in the graph
(see also Fig.~\ref{fig:cliquegraph} in the Appendix~\ref{a:suppfig}).
In Fig.~\ref{fig:gbs_rand_search} we show the outcomes of a numerical experiment
where a GBS device has been programmed to sample from the Hafnian of  $\Omega A \Omega$, with $\Omega$ as in Eq.~\eqref{e:weightrule}.
For simplicity, we choose $\alpha=1$ in Eq.~\eqref{e:weightrule}, although performance
can be slightly improved with optimized values of $\alpha$.
On the other hand, the parameter $c$ does not play any role in the post-selected data,
but it does change the overall probability of
getting samples of size $N=8$. For comparison, we have also studied
a purely classical random search,
where each data is a uniform random subgraph with $N$ vertices. We observe
only three cliques over $10^5$
samples.
On the other hand, as shown in Fig.~\ref{fig:gbs_rand_search}, GBS is able to produce roughly $300$ cliques directly
from sampling, without any classical post-processing.
This indicates that the GBS distribution is indeed favouring cliques with large weights, as intended.


Post-selecting on the number of detector clicks is an unwise strategy when employing real GBS devices because it disregards otherwise useful samples. Moreover, the size of the maximum weighted clique is generally unknown. Instead, we can generate cliques from every sample by employing the shrinking strategy discussed in Section \ref{s:gbsclassical}.

\begin{figure}[t]
	\centering
	\includegraphics[width=0.9\linewidth]{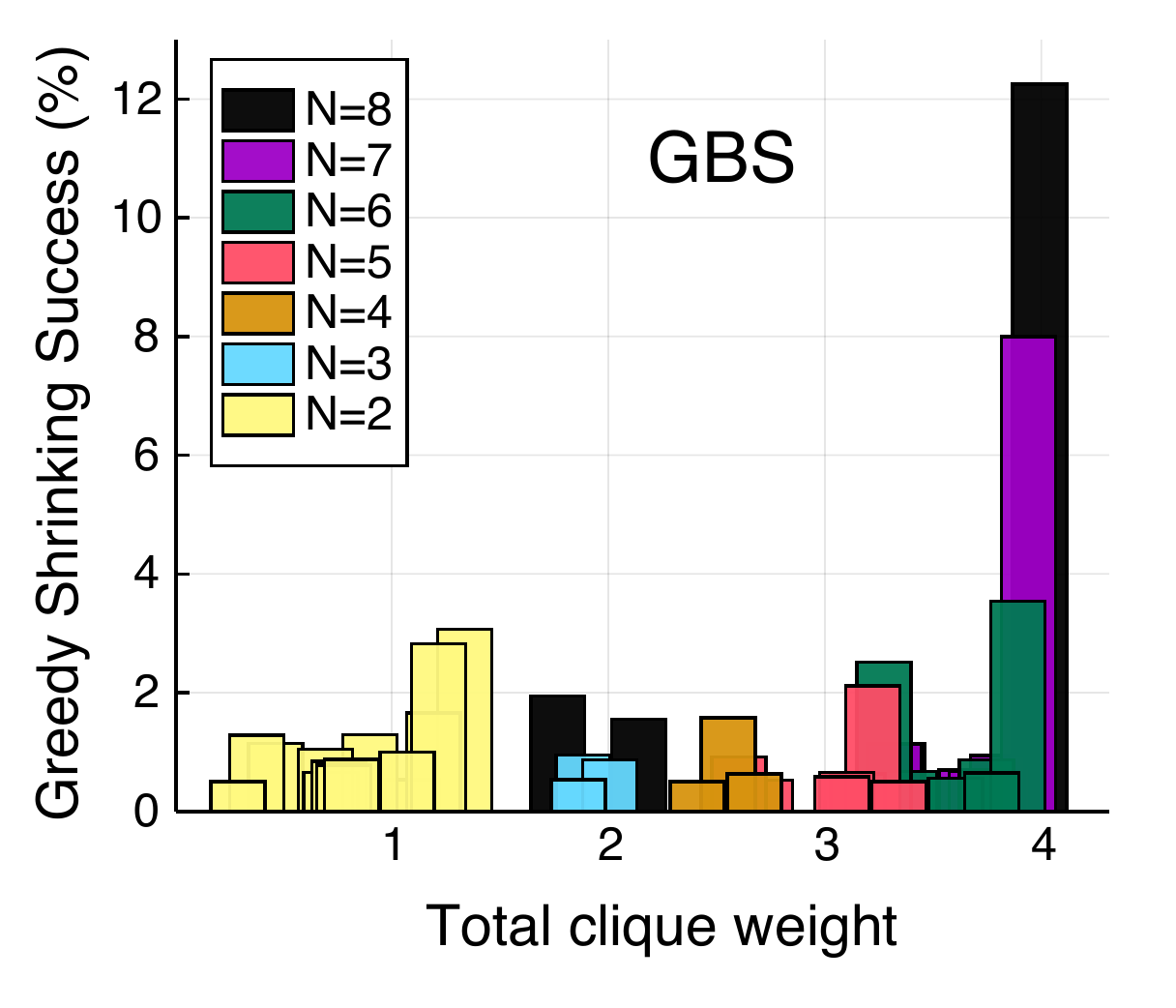}
	\caption{ {\bf Greedy shrinking success rate. }
		Success rate in finding cliques of different sizes ($N=2,\dots,N_{\rm max}$),
		when the max clique has size $N_{\rm max}=8$, as a function of the
		total clique weight $\sum_{j\in C} w_j$.  We used greedy shrinking over $10^4$
		GBS samples, ignoring trivial zero photon outcomes. Outcomes with low ($<0.5\%$)
		success rate are not shown.
	}
	\label{fig:greedy_shrink_wgbs}
\end{figure}

In Fig.~\ref{fig:greedy_shrink_wgbs} we study the performance of greedy shrinking with
GBS data. These data consist of $10^4$ samples
obtained from an exact numerical sampling algorithm \cite{gupt2018classical}.
Each sample corresponds to a subgraph and, unlike Fig.~\ref{fig:gbs_rand_search},
here any subgraph size is considered.
These results show that with GBS and greedy shrinking --
a simple classical post-processing
heuristic -- it is possible to obtain the maximum weighted clique with sufficiently high probability.
Indeed, the histogram in Fig.~\ref{fig:greedy_shrink_wgbs} has a sharp peak corresponding
to the clique of maximum size $N=8$ and maximum weight $\approx 3.99$. The success rate in sampling from the max weighted clique is
$\approx 12\%$ and the overall sampling rate for $N=8$ cliques is $\approx 19\%$.
Greedy Shrinking with purely classical random data is shown
in the Supplementary Fig.~\ref{fig:greedy_shrink_unif}.
Although the classical distribution is chosen to have
the same mean and variance as the GBS distribution, its performance is considerably worse: the maximum weighted clique is obtained only $1\%$ of the time, compared to $12\%$ for GBS.
This shows that GBS with greedy shrinking is already able to find the maximum weight clique of the graph after only a few repetitions.

\begin{figure}[t]
	\centering
	\includegraphics[width=0.9\linewidth]{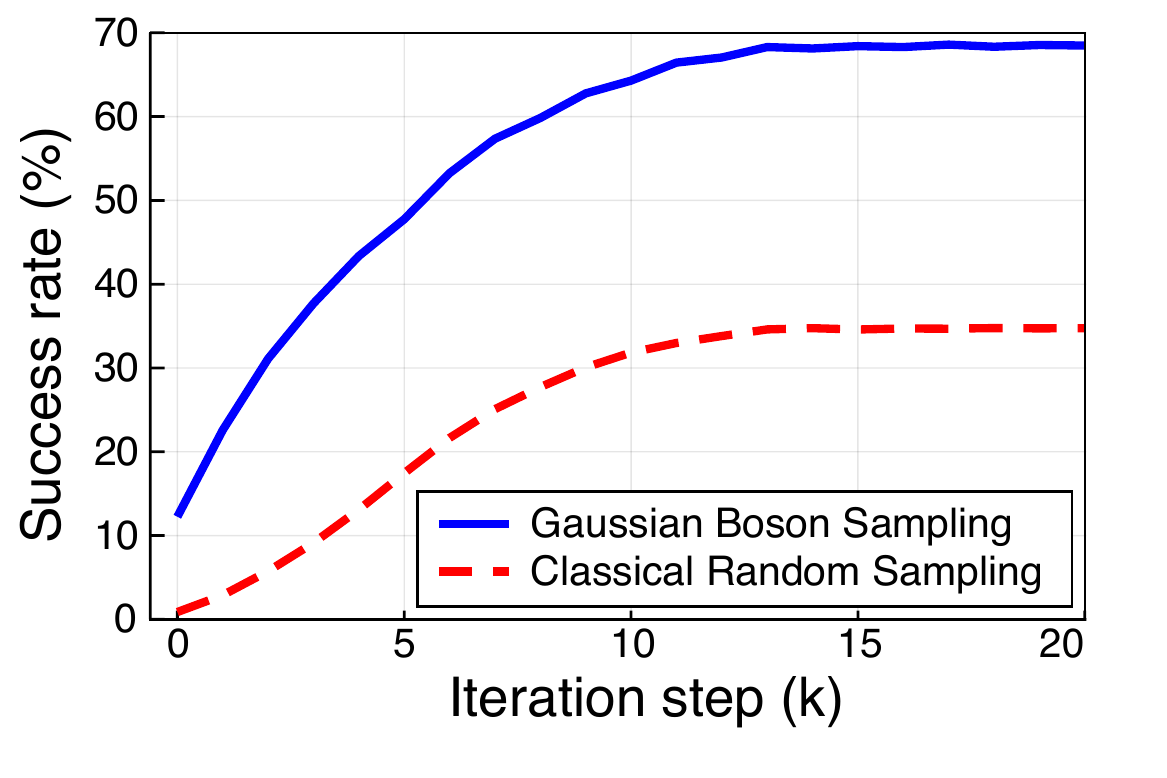}
	\caption{  {\bf GBS vs. classical success rate.}
		Success rate in finding the maximum weighted clique after greedy shrinking and
		$k$ expansion steps with local search. Samples are generated from either
		GBS or a purely classical approach.
		GBS maintains a significantly higher success rate over all iteration steps.
	}
	\label{fig:shrinkexpand_maxwcliq_simple}
\end{figure}

Finally, we study how the cliques obtained from GBS with greedy shrinking can be enlarged or
improved via local search.
Fig.~\ref{fig:shrinkexpand_maxwcliq_simple} shows the performance of the hybrid GBS shrinking + local search algorithm, compared to a classical strategy. The results indicate that GBS not only provides better initial estimates after greedy shrinking (zero iteration steps), but it maintains a significant margin
compared to classical strategies as the number of steps is increased. After $k=8$
local expansion steps, the
probability of finding the maximum weighted clique is as high as $60\%$, while the classical strategy has a considerably smaller success rate of $<30\%$. After many steps, the success
rate saturates: using GBS the success rate gets close to $70\%$,
while for the purely classical approach it remains under approximately $35\%$.


The role of noise and squeezing is discussed in Appendix~\ref{a:suppfig}, where we show that
GBS success rate is not diminished by the effect of noise, provided that the
amount of squeezing is increased accordingly. Therefore,
GBS shrinking and its variant with local search are
robust against noise, maintaining a significant margin compared to purely
classical strategies.

%% file: conclusions.tex

We have shown that Gaussian Boson Sampling (GBS) can be employed to predict accurate molecular docking configurations, a central problem in pharmaceutical research and development. 
This is achieved by first mapping the docking problem to the task of finding large cliques in a vertex-weighted graph, then programming the GBS device to sample these cliques with high probability. This constitutes an example of the viability of near-term quantum photonic devices to tackle problems of practical interest. 

Established algorithms for obtaining molecular docking configurations exist, but a challenge arises in the context of industrial drug design where large numbers of candidate molecules must be screened against a drug target.
In this case, a fast method for predicting docking configurations is required.
In principle, photonic devices such as Gaussian Boson Samplers can operate at very high rates, and may potentially provide solutions in shorter timeframes.
Additionally, by sampling better random subgraphs, GBS serves as a technique to enhance the performance of classical algorithms because it increases the success rate of identifying large weighted cliques.
This property is relevant and applicable in any context where identifying clusters in graphs is important, beyond applications in molecular docking.


More broadly, our results establish a connection between seemingly disparate physical systems: the statistical properties of photons interacting in a linear-optical network can encode information about the spatial configuration of molecules when they combine to form larger complexes.
In other words, we have found that when the interaction between fundamental particles is carefully engineered, they acquire collective properties that can be probed to perform useful tasks.
A complete understanding of the capabilities of emerging quantum technologies may thus require further exploration of systems that, even if incapable of universal quantum computation, can still be programmed to exhibit properties that can be harnessed for practical applications.

%% file: app_laplacian.tex
\section{Technical details}\label{a:technical} 

\subsection{ Pure-state Gaussian Boson Sampling}
Consider $M$ single-mode 
squeezed states with squeezing parameter $r_j$ injected into an $M\times M$ interferometer 
described by the unitary matrix $U$. For a pure Gaussian state, the matrix $\mathcal A$ 
entering in Eq.~\eqref{e:GBS} can be decomposed as $\mathcal A = B\oplus B^*$, where $B=U\bigoplus_{j=1}^M \tanh(r_j) U^T$ \cite{hamilton2017gaussian}. 
Therefore, the GBS device can be programmed with any matrix $B$ whose spectrum is 
contained in $[0,1]$. The adjacency matrix $A$ of a graph normally does not have this 
spectral property. However, it can always be rescaled as $A=cB + d\id$ where 
$B$ has the desired spectrum and $c,d$ are suitable rescaling constants. Using 
properties of the Hafnian, namely ${\rm Haf}(A) = c^N {\rm Haf}(B)$ and ${\rm Haf}(B\oplus B^*) = 
|{\rm Haf}(B)|^2$, it was found that a GBS device can be 
programmed to sample from a distribution $p(S)\propto \frac{|{\rm Haf}(A_S)|^2}{c^N}$ \cite{bradler2018gaussian}.

\bigskip

\subsection{ Weighted Gaussian Boson Sampling }
We now introduce some technical results to show the properties of weighted 
Gaussian Boson Sampling.  
\begin{lemma}
Let $B$ be defined as in Eq.~\eqref{e:Bmat}, where $\Omega$ 
is diagonal with non-zero diagonal elements $c\omega_j$ for some $c$. 
Then, $B$ has spectrum in $[0,c]$ if $	2 \max_j d_j\omega_j^{2}\le c^{-1}$.
\end{lemma}
\noindent {\it Proof.}~~
The proof closely follows that of Lemma 1.7 in \cite{chung1997spectral}. 
Since the Laplacian $D-A$ is positive semidefinite, so is $B$. 
Let $f$ be any vector, then 
\begin{align}
	\sup_f \frac{f^TBf}{f^Tf} &= 
	\sup_f \frac{f^T\Omega D^{1/2}\tilde{L}D^{1/2}\Omega f}{f^Tf} =\cr&= 
	\sup_g \frac{g^T\tilde{L}g}{g^Tg}\frac{g^Tg}{g^TD^{-1/2}\Omega^{-2}D^{-1/2}g} \le \cr &\le 
	\sup_g  \frac{2 g^Tg}{g^TD^{-1/2}\Omega^{-2}D^{-1/2}g} = \cr & =
	\sup_f  \frac{2 f^T\Omega D\Omega f}{f^Tf} \le 
	\max_j {2 c^2 \omega_j^2 d_j}
	~,
\end{align}
where $g=D^{1/2}\Omega f$ and where we used that the spectrum 
of the normalized Laplacian $\tilde L$ is contained in $[0,2]$. 
\hfill\qedsymbol

\bigskip
The main result of this section is the following decoupling theorem
\begin{theorem}
	Let $B$ be the matrix defined in \eqref{e:Bmat}. Then
\begin{equation}
	{\rm Haf}(B) = \det(\Omega) \; {\rm Haf}(A)~.
	\label{e:wid}
\end{equation}
\end{theorem}
\noindent \textit{Proof.}~~
The proof is based on the following expansion 
\begin{align}
	{\rm Haf} (B) &\stackrel{(a)}{=} \sum_{M \in \rm PMP} \prod_{(ij)\in M } (B)_{ij} =\cr &
	\stackrel{(b)}{=} \sum_{M \in \rm PMP} \prod_{(ij)\in M } A_{ij} 
	\prod_{(ij)\in M } (\omega_i\omega_j)
	=\cr & \stackrel{(c)}{=} \prod_k (\omega_k) \sum_{M \in \rm PMP} \prod_{(ij)\in M } A_{ij}~,
\end{align}
where in (a) we use the definition of the Hafnian where PMP is the set of perfect matchings.
Equality (b) follows from the definition of $B$, being the Hafnian of a matrix 
independent of its diagonal elements. In (b) each $M$ contributes to the sum only 
if $A_{ij}\neq0$ for all $(ij)\in M$. In the latter case the product 
$	\prod_{(ij)\in M } (\omega_i\omega_j)$ is a product over all possible $\omega_j$, 
as each vertex is visited in $M$ only one time. 
In (c) we use  the fact that the latter product is independent of $M$. 
\hfill\qedsymbol

\bigskip

\subsection{ Biasing the number of detections }\label{a:bias}
We discuss the role of parameter $c$ in biasing the 
average output size. Consider a single-mode state with squeezing parameter $r$. 
Being pure, the $\mathcal A$ matrix is written as $\mathcal A=B\oplus B^*$ and,
for a single mode, $B=\tanh(r)$.
For maximum squeezing $r_{\rm max}$ we find that $B$ can take any value in 
$[0,c]$ with $c= \tanh(r_{\rm max})$. The resulting average photon number is then 
$\langle N\rangle = \sinh(r)^2 = \frac{c^2}{1-c^2}$ and the variance is 
$\Delta N^2 \propto{\langle N\rangle(1+\langle N\rangle )}$. For multiple modes 
the expressions are similar, though $B$ is a matrix and 
$\langle N\rangle = \Tr[\frac{B^2}{\mathbb I-B^2}]$,
so the normalization factor can be tuned to provide a higher rate to subgraphs 
of different sizes $N$. Although the maximum clique size is not known a priori, an estimate, 
e.g. based on random graphs \cite{bollobas1976cliques}, is normally enough as the large
variance $\Delta N^2 \approx {\langle N\rangle(1+\langle N\rangle )}$ assures that 
different sizes are sampled with sufficiently high rate. 

Gaussian boson sampling using click detectors yields a discrete 
probability distribution over subsets $S_N$ of $\{1,\dots,M\}$ of dimension $N$. 
We write $i\in S_N$ if the $i$th detector ``clicks'' and  $i\notin S_N$ otherwise.
The resulting probability distribution is \cite{quesada2018gaussian}
\begin{equation}
	p(S_N) = \Tr\left[\prod_{i\in S_N} P_1^i \prod_{i\notin S_N} P_0^i \;\rho\right]~,
	\label{e:torontonianP}
\end{equation}
where $P_0^{i} = \vert0_i\rangle\langle{0_i}\vert$ is the projection into the 
zero photon state and $P_1^i=\openone-P_0^i$. 
The average number of clicks $N$ is then 
\begin{equation}
	N   = \sum_{j=1}^M \Tr[P_1^j \rho] = M-\sum_{j=1}^M \langle0\vert\rho_j\vert{0}\rangle~,
\end{equation}
where $\rho_j$ is the reduced state on mode $j$. 
Using the fidelity formula for Gaussian states \cite{banchi2015quantum} we then get
\begin{equation}
	N[\sigma] 
	= M-\sum_{j=1}^M \frac{1}{\sqrt{\det[\sigma_{j} + \openone /2]}}~,
\end{equation}
where $\sigma_j$ is the reduced ($2\times 2$) covariance matrix for mode $j$. 
The above equation can be solved to bias the number of clicks. When 
the covariance matrix $\sigma$ depends on the normalization factor $c$, we can 
use a simple line search algorithms to tune $c$ such that $N[\sigma(c)]$ is equal to 
the desired value.

\bigskip

\bigskip

\subsection{Post-selection } \label{a:postselection}
Sampling from Eq.~\eqref{e:torontonianP} requires the 
calculation of all $p(S_N)$ for $N=1,\dots,M$. There are exponentially many of 
these probabilities $\mathcal O(2^M)$. However, if we are interested in samples 
of a fixed size $N$, then the number of $p(S_N)$ with fixed $N$ is 
$\mathcal O(\binom{M}{N}) \approx \mathcal O(M^N)$. Each probability requires 
the evaluation of $\mathcal O(2^N)$ determinants, 
so the complexity is still exponential as a function of $N$ \cite{quesada2018gaussian}.
However, focusing on postselection with a 
certain size $N$ reduces the complexity of brute force approaches from exponential 
to polynomial, although the degree of this polynomial increases with $N$.

\bigskip
\subsection{ Selecting parameter $\alpha$}
For a complete graph with $2n$ vertices
the Hafnian is $h_{n}= \frac{2n!}{n!2^n}$. The largest Hafnian for non-complete graphs
is obtained by removing an edge from the complete graph. The Hafnian is then
$\frac{2n-2}{2n-1}h_n$, so this non-optimal graph is penalized by a factor 
$\frac{2n-2}{2n-1} \simeq 1-\frac1{2n}$. A possible choice for $\alpha$ is to 
avoid a counterbalance of this term, so $1+\alpha w_{\rm tot} < \frac{2n-1}{2n-2}$.
Nonetheless, we have numerically observed that, at least for sparse graphs, the 
parameter $\alpha$ does not have to be carefully chosen, and different values of
$\alpha$ provide the expected enhancement for the max weighted clique problem. 

\bigskip

%% file: app_labdistgraph.tex
\section{Graph representations of molecular interactions}\label{a:graphrepr}

In this section, we use $\mathbb{L}$ to denote the set of all labels corresponding to the individual pharmacophore point types
and $\kappa$ to denote the potential function $\kappa: \mathbb{L}\times \mathbb{L} \rightarrow \mathbb{R}$ that assigns an interaction strength to each pair of labels from $\mathbb{L}$.

\subsection{Labeled distance graph}

\begin{definition}{\textit{Labeled distance graph.}}
\label{def:ldg}
Let $S$ be a set of points in three dimensional space $S = \big\{(x_i, y_i, z_i) \,|\, i \in I \big\}$ for a given index set $I$ heuristically selecting pharmacophore points of a component (either ligand or the binding site) involved in the binding complex.
Then \textit{labeled distance graph} $G_S$ is defined as $G_S = (V_S, E_S, \omega_S, \alpha_S)$ where
\beq
V_S = \big\{ v_i \,|\, \boldsymbol{p_i} \in S \big\}
\eeq
is the set of vertices,
\beq
E_S = \big\{ (v_i, v_j) \,|\, \boldsymbol{p_i,p_j} \in S, i < j \big\}
\eeq
is the set of edges,
\beq
\omega_S((v_i, v_j)) = || \, \boldsymbol{p_i} - \boldsymbol{p_j}\, ||
\eeq
is the weighting function of the edges and
\beq
\alpha_S: V_S \rightarrow \mathbb{L}
\eeq
is a function assigning a pharmacophore point type to each vertex.
\end{definition}
\begin{remark}
Any labeled distance-graph is a complete graph with $I$ vertices.
\end{remark}

\subsection{Binding interaction graph}

Let $G_L = (V_L, E_L, \omega_L, \alpha_L)$ be a ligand labeled distance graph and $G_B = (V_B, E_B, \omega_B, \alpha_B)$ labeled distance graph for the binding site.
Any pair of vertices $(l_i, b_i) \in V_L \times V_B$ is then called a \textit{contact} between $G_B$ and $G_L$.

\begin{definition}{\textit{$\tau$ flexible contact pair.}}
\label{def:tauflex}
Let $c_i = (l_i, b_i)$ and $c_j = (l_j, b_j)$ be contacts between $G_L = (V_L, E_L, \omega_L, \alpha_L)$ and $G_B = (V_B, E_B, \omega_B, \alpha_B)$.
Then $(c_i , c_j)$ is a \textit{$\tau$ flexible} contact pair between $G_L$ and $G_B$ if and only if $|\, \omega_L(l_i, l_j) - \omega_B(b_i, b_j)\, | \leq \tau + 2\epsilon $, where $\tau$ is the flexibility constant and $\epsilon$ is the interaction cutoff distance.
\end{definition}
\begin{remark}
    Mutual $\tau$ flexibility of contact pairs is a reflexive and symmetric relation, but not necessarily transitive.
\end{remark}

For multiple contact pairs to be realized in the binding pose they have to not violate each other's geometric constraints and hence be pairwise $\tau$ flexible . In the following graph representation, this corresponds to a clique:

\begin{definition}{\textit{Binding interaction graph.}}
\label{def:big}
    Let $G_L = (V_L, E_L, \omega_L, \alpha_L)$ be a labeled distance-graph for a given ligand and $G_B = (V_B, E_B, \omega_B, \alpha_B)$ a labeled distance-graph for a given binding site.
    The corresponding binding interaction graph $I_{L,B}$ is defined as
    \beq
    I_{L,B} = (\mathcal{V}, \mathcal{E}, \kappa, \tau, \epsilon),
    \eeq
    where vertex set $\mathcal V$ is the set of the pairs over vertex-sets of $G_L$ and $G_B$
    \beq
    \mathcal{V} = V_L \times V_B,
    \eeq
    and $\tau,\epsilon \in \mathbb{R^{+}}$ are the flexibility threshold constant and interaction cutoff distance. Then
    \beq
    \mathcal{E} \subseteq \Big\{ \big((v_{l_1}, v_{b_1}),(v_{l_2}, v_{b_2}) \big) \, | \, v_{l_1},v_{l_2} \in V_L, v_{b_1},v_{b_2} \in V_B \big\}
    \eeq
        is a maximal set of $\tau$ flexible contact pairs between $G_L$ and $G_B$ and $\Omega: \mathcal{V} \rightarrow \mathbb{R}$ is an vertex-weighting function defined as
    \beq
    \Omega\big((v_l, v_b)\big) = \kappa\big(\alpha_L(v_l), \alpha_B(v_b)\big),
    \eeq
    which encodes the interaction strength between pharmacophore points corresponding to vertices $v_l$ and $v_b$.
\end{definition}
\begin{remark}
    Most favourable binding pose of ligand described by labeled distance graph $G_L$ and binding site described by labeled distance graph $G_B$ corresponds to the heaviest vertex-weighted clique of binding interaction graph $I_{L,B}$.
\end{remark}

%% file: app_data.tex
\section{TACE-AS complex }\label{a:data}


The binding interaction graph used in numerical simulations is constructed as follows. 
First, using the software package \texttt{rdkit} \cite{rdkit} all the pharmacophore points on ligand and receptor are extracted.
This results in 11 pharmacophore points on the AS ligand and 243 points on the TACE receptor.
These sizes are not too large for future GBS devices, but the classical
simulation of GBS is intractable for such large problem instances \cite{gupt2018classical}. 
Therefore, to enable numerical simulations, we subselect
pharmacophore points based on the true binding pose of AS and TACE according to
the following two criteria:
\begin{enumerate}
    \item Select pairs of pharmacophore points on ligand and receptor that are within $4$\r{A} distance of each other.
    \item From these pairs select the ones whose label pairs are either hydrogen donor/acceptor, hydrophobe/hydrophobe, negative/positive charge, aromatic/aromatic.
\end{enumerate}
Note that in a realistic scenario the true binding pose would be unknown.
However, a similar set of points could be obtained based on knowledge commonly employed in drug discovery.
For example, ligand pharmacophore points can be heuristically selected and prior knowledge of the binding site location drastically reduces the number of pharmacophore points on the receptor.
To reduce the number of receptor pharmacophore points even further, one could use a sliding window to study different sections of the binding site in isolation.
Nevertheless, these reduction techniques will be unnecessary when physical GBS devices are built 
with enough modes. 
In the case of the TACE-AS complex, we subselect 4 points on the ligand and 6 points on the receptor and create two labelled distance graphs as illustrated in Fig.~\ref{fig:labelled_distance_graph}.

Using PDBbind \cite{liu2014pdb,wang2005pdbbind,wang2004pdbbind}, a curated dataset of protein-ligand interactions, we derive a knowledge-based pharmacophore potential.
For all protein-ligand interactions, the pharmacophores on ligand and binding site are extracted with \texttt{rdkit} \cite{rdkit} and all pairwise distances are accumulated in a single histogram.
Subsequently, the Drugscore potential \cite{gohlke2000predicting,gohlke2000knowledge} for the six pharmacophore types (negative/positive charge, hydrogen donor/acceptor, hydrophobe and aromatic ring) is computed from the histogram as outlined in Ref.~\cite{mooij2005general}.
The Drugscore potential yields values in the interval $[0,1]$ whereby favourable interactions are close to $0$.
Since we want to encode the correct binding pose in a maximum weighted clique, we reflect the resulting potential,
\begin{equation}
    \label{eq:reflection}
    P_\text{refl}(i,j) = \max (P) - \min (P) - P_\text{orig}(i, j),
\end{equation}
such that large values in the potential encode desirable interactions.
\begin{table*}[t]
\begin{tabular}{l | m{2cm} | m{2cm} | m{2cm} | m{2cm} | m{2cm} | m{2cm}}
Pharmacophore type     & Negative charge  & Positive charge & Hydrogen-bond donor & Hydrogen-bond acceptor & Hydrophobe & Aromatic \\ \hline
Negative charge        & 0.2953           &                 &                     &                        &            &          \\ \hline
Positive charge        & 0.6459           & 0.1596          &                     &                        &            &          \\ \hline
Hydrogen-bond donor    & 0.7114           & 0.4781          & 0.5244              &                        &            &          \\ \hline
Hydrogen-bond acceptor & 0.6450           & 0.7029          & 0.6686              & 0.5478                 &            &          \\ \hline
Hydrophobe             & 0.1802           & 0.0679          & 0.1453              & 0.2317                 & 0.0504     &          \\ \hline
Aromatic               & 0.0              & 0.1555          & 0.1091              & 0.0770                 & 0.0795     & 0.1943   \\ \hline
\end{tabular}
\caption{ {\bf Knowledge-based pharmacophore potential}. Data is derived from the PDBbind dataset from 2015 \cite{liu2014pdb, wang2005pdbbind,wang2004pdbbind}. The matrix is lower-diagonal since any potential function is symmetric.}
\label{t:potential}
\end{table*}
The resulting knowledge-based potential is shown in Table~\ref{t:potential}.


\bigskip

%% file: app_suppfig.tex
\section{Supplementary figures}\label{a:suppfig}

\begin{figure*}[ht]
	\centering
	\includegraphics[width=0.95\linewidth]{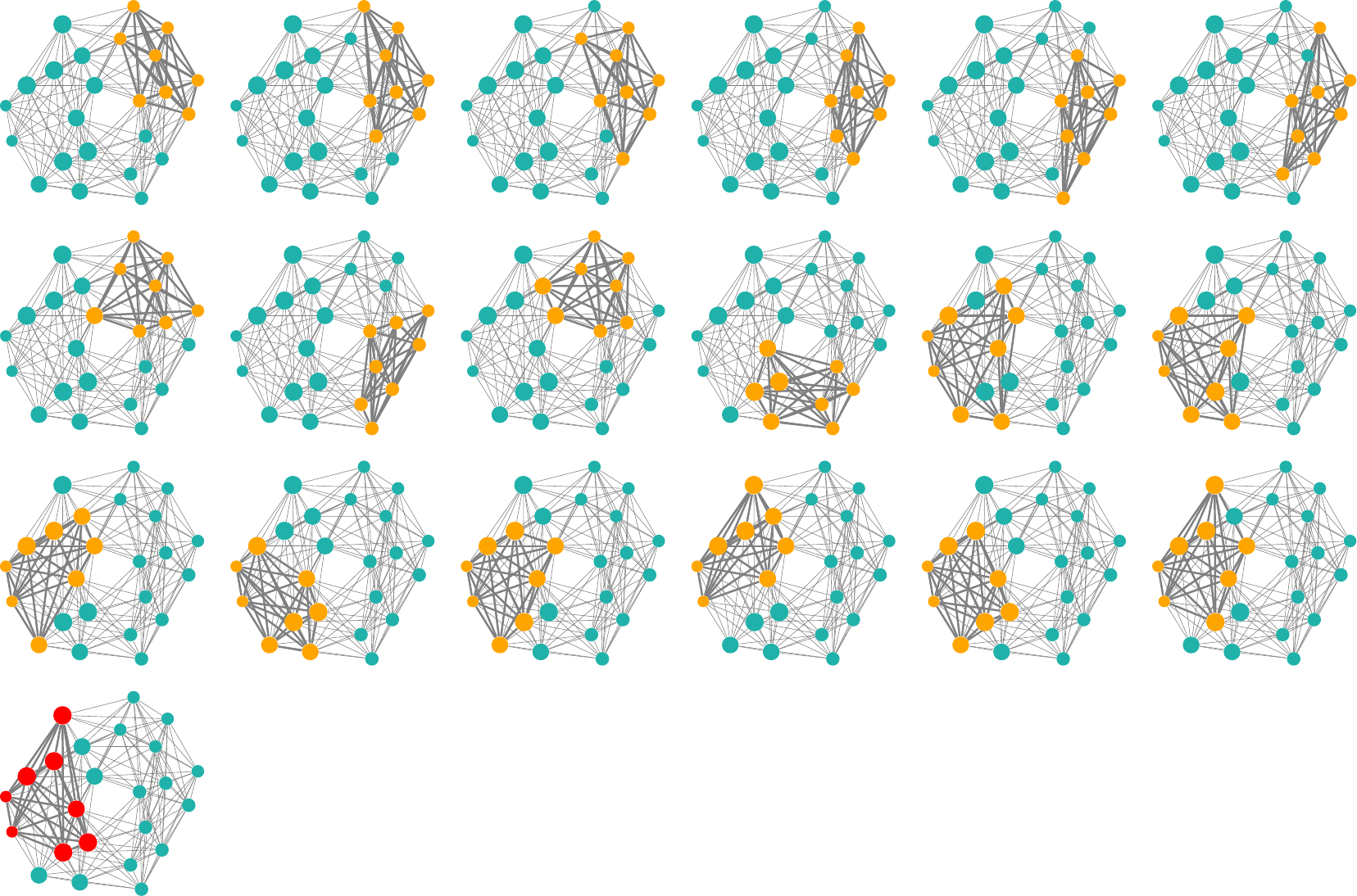}
	\caption{ {\bf Position of all the maximum cliques.} 
		We focus on the TACE-AS graph, where cliques are
        shown with orange notes, darker edges. 
		The diameter of each vertex is proportional to its weight.
		The cliques are ordered from low to high total weight, starting from
		the top-left until the bottom-right order. The heaviest clique is shown 
		in red in the last graph. 
	}
	\label{fig:cliquegraph}
\end{figure*}

In Fig.~\ref{fig:cliquegraph} we show the position of all maximum cliques (of size $N=8$) 
inside the graph, ordered from lightest to heaviest total weight. The figure shows that 
there are two main clusters in the graph: the top right cluster, generally with light weights, 
and the bottom left cluster with heavy weights. There are also a couple of intermediate cliques 
where these two clusters are mixed. 
The maximum weighted clique is shown in the bottom graph, where from node diameter 
we observe that it is composed by a heavy six-vertex core and two light vertices. 
Comparison with Fig.~\ref{fig:gbs_rand_search} shows that all lightweight cliques 
have a low occurrence rate  in a carefully programmed GBS device.

\begin{figure}[h]
	\centering
\includegraphics[width=0.9\linewidth]{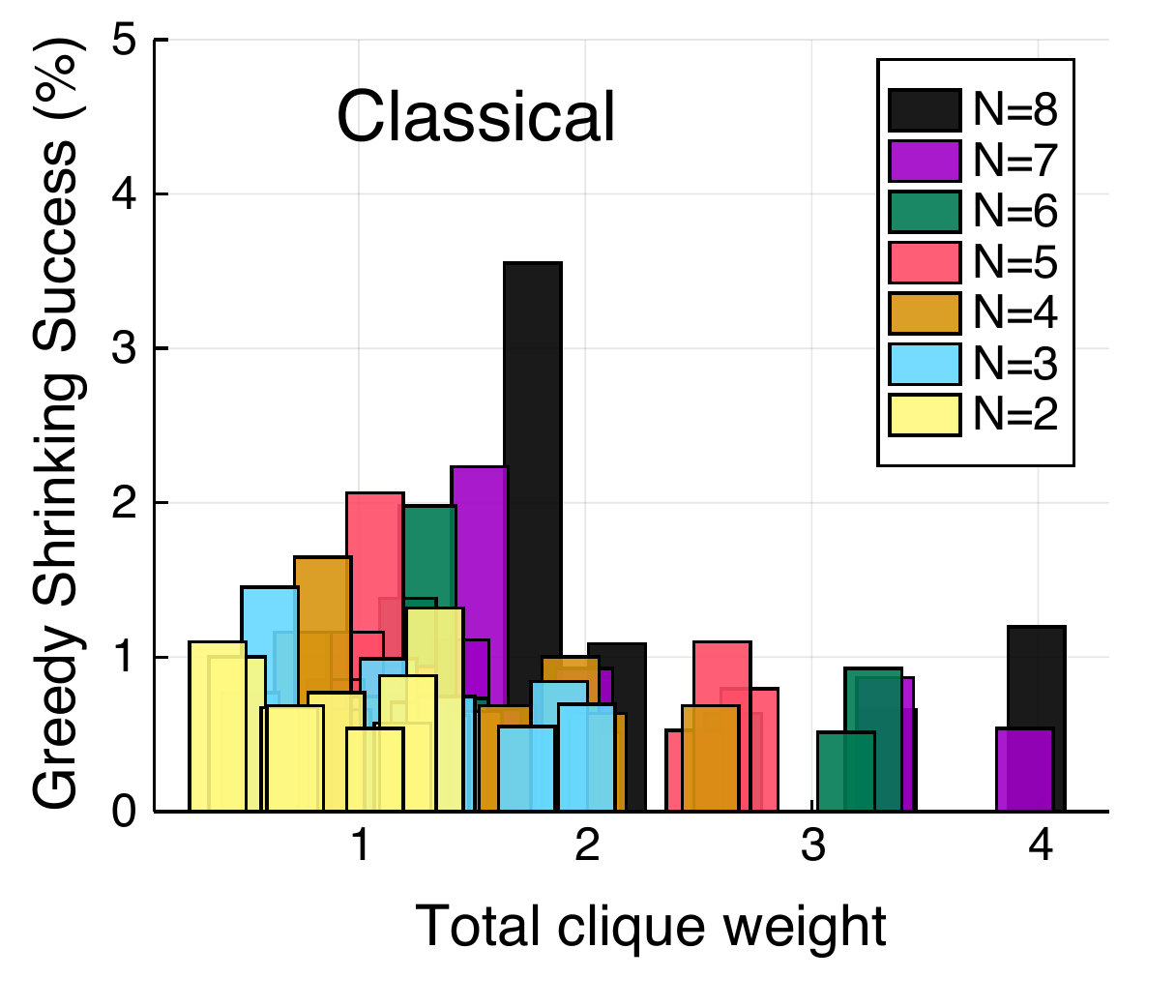}
	\caption{ 
		{\bf Classical greedy shrinking success rate.}	
		Success rate in finding cliques of different sizes ($N=2,\dots,N_{\rm max}$),
		when the max clique has size $N_{\rm max}=8$, as a function of the 
		total clique weight $\sum_{j\in C} w_j$.  We used greedy shrinking over $10^4$ 
		classical random samples. For fair comparison with GBS, classical samples where 
		generated by first sampling a size $N$ with same average and variance as GBS, and 
		then selecting a random subgraph with $N$ vertices. 
	}
	\label{fig:greedy_shrink_unif}
\end{figure}

In Fig.~\ref{fig:greedy_shrink_unif} we show the output of Greedy Shrinking
with purely classical random data.  For a fair comparison with the
GBS-based approach shown in Fig.~\ref{fig:greedy_shrink_wgbs},
the classical data are generated as follows: we first sample a subgraph size
$N$ from a normal distribution with the same mean $\langle N\rangle$ and
variance $\Delta N^2$ as the GBS distribution, then uniformly generate a random
subgraph with size $N$. Although the resulting distribution 
has the same mean and variance as the GBS distribution, by comparing 
Fig.~\ref{fig:greedy_shrink_unif} and  Fig.~\ref{fig:greedy_shrink_wgbs}, we see that
its performance is considerably worse: the maximum weighted clique is obtained
only $1\%$ of the time, compared to $12\%$ for GBS.

\begin{figure}[t]
	\centering
	\includegraphics[width=0.9\linewidth]{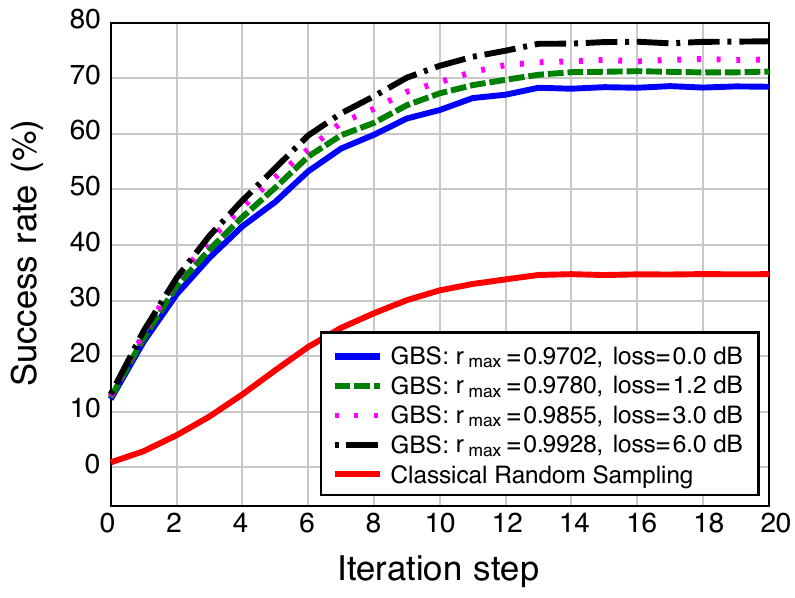}
	\caption{ {\bf Success rate vs. squeezing and noise.} 
		Success rate in finding the maximum weighted clique after greedy shrinking and local 
		search. GBS is compared to a purely classical approach. 
		For GBS, different values of squeezing and noise are considered. 
	}
	\label{fig:shrinkexpand_maxwcliq_noise}
\end{figure}

In Fig.~\ref{fig:shrinkexpand_maxwcliq_noise} we study the effect of noise and
squeezing. The value $r_{\rm max}= 0.9702$ corresponds to an average number of
detector clicks $\langle N\rangle \simeq 8$. In the lossy case, for a fair comparison,
we have increased the squeezing to $r_{\rm max}= 0.9780$ in order to maintain
the same average  $\langle N\rangle \simeq 8$ and have, accordingly, samples of the
same average size. 
As Fig.~\ref{fig:shrinkexpand_maxwcliq_noise} shows, the success rate is not diminished 
by the effect of noise, provided that the amount of squeezing is increased accordingly. 
As a matter of fact, the noisy version with larger squeezing displays a similar success 
rate after greedy shrinking (iteration 0). As the iterations increase, the success rate 
of both noisy and noiseless GBS maintain a significant margin compared to the purely classical
strategy. The slightly better performance of the noisy case is due to the larger squeezing 
that changes the shape of the photon distribution,  while keeping comparable photon 
averages with the noiseless case. 
This analysis shows that both GBS shrinking and its variant with local search are robust 
against noise, maintaining a significant margin compared to purely classical strategies.